\documentclass[usenatbib,useAMS]{mn2e}

\usepackage{amssymb}
\usepackage{subfigure}
\usepackage{longtable,lscape}
\usepackage{threeparttable}
\usepackage{verbatim}
\usepackage{rotating}
\usepackage{color}

\newcommand{\comments}[1]{}
\newcommand{\tco}{\mbox{$^{13}$CO(3-2)}}
\newcommand\nhp{\mbox{N$_2$H$^+$}}

\newcommand\hcop{HCO$^+$}

\newcommand\hnc{HNC}
\newcommand\hnco{HNCO}
\newcommand\ch{C$_2$H}
\newcommand\htcop{H$^{13}$CO$^+$}
\newcommand\hctn{HC$_3$N}
\newcommand\hntc{HN$^{13}$C}
\newcommand\tcs{$^{13}$CS}

\newcommand\kms{km s$^{-1}$}

\newcommand{\mum}{$\mu$m}

\bibpunct{(}{)}{;}{a}{}{,}

\defcitealias{Contreras-2013b}{Paper I}

\title[Fragmentation in filamentary molecular clouds]{Fragmentation in filamentary molecular clouds}

\author[Contreras et al.]{Yanett Contreras$^{1,2}$\thanks{E-mail: ycontreras@strw.leidenuniv.nl}, Guido Garay$^{2}$, Jill M. Rathborne$^{1}$ and Patricio Sanhueza$^{3}$\\
  $^{1}$CSIRO Astronomy and Space Science, PO Box 76, Epping NSW 1710, Australia\\
  $^{2}$Departamento de Astronom{\'i}a, Universidad de Chile, Casilla
  36-D, Santiago, Chile\\
  $^{3}$National Astronomical Observatory of Japan, 2-21-1 Osawa, Mitaka, Tokyo 181-8588, Japan}

\begin{document}

\date{Accepted 2015 November 26.  Received 2015 November 25; in original form 2015 May 15}

\maketitle

\label{firstpage}

\begin{abstract}

Recent surveys of dust continuum emission at sub-mm wavelengths 
have shown that filamentary molecular clouds are ubiquitous along the 
Galactic plane. These structures are inhomogeneous, with over-densities that are sometimes associated with infrared emission and active of star formation. To investigate the 
connection between filaments and star formation, requires an understanding of the processes that lead to the fragmentation of filaments and  a determination of the 
physical properties of the over-densities (clumps). In this paper, we present a multi-wavelength study of five filamentary molecular clouds, containing several clumps in 
different evolutionary stages of star formation. We analyse the fragmentation of the filaments and derive the physical properties of their clumps. We find 
that the clumps in all filaments have a characteristic spacing consistent with 
the prediction of the `sausage' instability theory, regardless of the complex morphology of the filaments or their evolutionary stage. We also 
find that most clumps have sufficient mass and density to form high-mass stars, supporting the idea that high-mass stars and clusters form within filaments. 
  
\end{abstract}


\newpage
\section{Introduction}

One striking result of recent dust continuum Galactic Plane surveys is the 
ubiquitous presence of filamentary structures (e. g., Apex Telescope Large Area Survey (ATLASGAL), \citealt{Schuller}; 
Bolocam, \citealt{bolocam}; Herschel infrared Galactic Plane Survey (HiGAL), \citealt{hershel}). Their study has recently gained 
momentum, allowing us to characterize their overall properties and to better understand their formation. 
\citep[e.g.,][]{teixeira, myers-2009, Andre-2010, Arzoumanian, Hernandez-2011, 
Myers-2011, peretto-2012}. Filamentary molecular clouds have high aspect ratios (typically larger than 3:1), are often infrared (IR) dark, and harbour over-densities or star-forming clumps along their length. 

The processes that form clumps within filamentary molecular clouds are still not well 
understood. Theoretical work has shown that filaments can be subject to gravitational instabilities, and they should fragment 
into clumps which are located at a characteristic spacing, given by the wavelength of 
the fastest growing unstable mode \citep{chandrasekhar-1953, nagasawa-1987, 
inutsuka-1992, nakamura-1993, tomisaka-1995}. Perturbations that are greater 
than the characteristic wavelength will become rapidly unstable 
and clumps (over densities) will be formed within the filament. This preferred spacing 
depends on the intrinsic characteristics of the filament, the type of 
fluid (e.g., incompressible, isothermal) and on the shape and strength of any associated
magnetic field. 

\citet{fiege-pudritz-2000} modelled the fragmentation of filamentary 
molecular clouds with a helical magnetic field. They found that filaments that are constrained by these types of magnetic fields are susceptible to both gravitationally driven instabilities and 
magneto-hydrodynamic (MHD) driven instabilities. One kind of MHD instability that predicts the formation of regularly 
spaced clumps in a filament is the fluid or `sausage' instability. This mechanism 
has recently been invoked to explain the regular spacing observed in some filamentary 
molecular clouds \citep[e.g.,][]{Jackson-2010, Miettinen-2010, Wang-2011}. 

Once the filament fragments and clumps are formed, some of the clumps will evolve, 
collapse, and form stars. IR emission associated with some filaments show evidence for heated and shocked gas. Since these are often produced by embedded protostellar objects and are indicators of active star-formation, recent work has proposed that filaments may be important for the process of star formation \citep[e.g.][]{Schisano-2014, Sanhueza-2010}. 
Moreover, some studies have suggested that filaments can also enhance accretion 
rates on to individual star-forming cores by material flowing from the filament to the clumps, and 
thus potentially facilitating the formation of high mass stars \citep{banerjee-2008, myers-2009}. 
To determine whether filamentary molecular clouds and their clumps have the ability to harbour high-mass stars, we have undertaken a detailed study of their global physical properties and their potential to form high-mass stars.

In this paper we characterize the fragmentation within five filamentary clouds and determine the physical properties and star-formation potential of their embedded clumps. The 
overall properties of these filaments were previously reported by Contreras et 
al. (2013) (\citetalias{Contreras-2013b} hereafter, see Table \ref{summ-frag1}). 

The five filaments are physically coherent structures in position-position-velocity space, shown via their molecular line emission from the low-density tracer $^{13}$CO (3-2). They were selected as contiguous emission features in the ATLASGAL 870 \mum~dust continuum emission survey and named after their brightest embedded clump using their ATLASGAL catalogue 
denomination \citep{Contreras-2013a}. For simplicity, hereafter we refer to each filament with a letter (A through E): `A: AGAL337.922-0.456' (also known as the `Nessie Nebula', \citet{Jackson-2010}). 
`B: AGAL337.406-0.402', `C: AGAL335.406-0.402', `D: AGAL332.294-0.094', and 
`E: AGAL332094-0421'.  
  
The filaments are associated with varying degrees of IR emission identified using the 
\textit{Spitzer} MIPS Inner Galactic Plane Survey \citep[MIPSGAL;][]{carey-2009} 
and in the Galactic Legacy Infrared Midplane Survey Extraodinaire 
\citep[GLIMPSE;][]{benjamin-2003} images. Indeed, it is often the case that within a single filament there can be regions that are IR dark, while others can be IR bright and clearly associated with active star formation (see Figures 1-5). Since the presence of IR emission is a signpost of star formation, the varying degree of star formation within these filaments will allow us to also investigate whether or not the separation between clumps change as the clumps evolve.
 
Moreover, these filaments have a lineal 
mass (mass per unit length) smaller than their critical mass (virial mass per 
unit length) suggesting that they are supported by an helical toroidal dominated 
magnetic field \citepalias{Contreras-2013b}. Thus, MHD driven instabilities, such as the `sausage' instability, may be responsible for the formation of the clumps within these filaments. 

Section 2 describes the dust continuum and molecular line observations made towards 
the filaments. Section 3 summarizes the physical properties of the embedded 
clumps. In Section 4 we discuss the star formation activity within the clumps determined from their IR and molecular line emission. 
In Section 5 we analyse the fragmentation of the filaments and discuss the potential of the 
clumps to form high-mass stars.

\section{The Data}

\subsection{ATLASGAL: 870 \mum~dust continuum emission}\label{sec:surveys}

We used the ATLASGAL
\citep[][]{Schuller} maps to measure the dust thermal emission at 
870 \mum~ towards each filament. The ATLASGAL survey was carried out using the Large Apex BPlometer CAmera (LABOCA) bolometer receiver \citep{siringo-2009} mounted at Atacama
Pathfinder EXperiment \citep[APEX;][]{guesten}, located in the Llano de Chajnantor, Chile. At this frequency the APEX beam size is 19$''$.2. The uncertainty in flux is estimated to be
lower than 15\% \citep{Schuller} and the pointing rms is $\sim$~4". The 1 $\sigma$ noise of the data is typically 50
mJy beam$^{-1}$.

\subsection{MALT90: 90 GHz dense gas emission}\label{sec:mat90-cores}

We use data from the Millimetre Astronomy Legacy 
Team 90 GHz survey \citep[MALT90;][]{Foster-2013, jackson-2013}  to determine the dynamical state of the clumps and to derive the properties of the 
dense gas within the clumps. The MALT90 survey covered 80 of the 101 clumps within these filaments. Hereafter, we will refer to these clumps as MALT90 clumps.

The MALT90 survey was conducted using the 8 GHz
wide Mopra Spectrometer (MOPS) at the Mopra 22-m radio telescope, simultaneously mapping 16 
spectral lines near 90 GHz (Table \ref{lines-malt90}). The maps have an angular and spectral resolution of 38" and \mbox{0.11~\kms}, respectively. The 
typical $1\sigma$ noise in the spectra is T$_A^* \sim$ 0.2~K per channel.
The pointing is accurate to 8" and the system gain is repeatable to within  30\% \citep{Foster-2013}.
The maps were reduced by the MALT90 team using an automated reduction
pipeline. The reduced and calibrated data products are available to the
community via Australian Telescope Online Archive, 
ATOA\footnote{http://atoa.atnf.csiro.au/malt90}. 

The transitions observed in this survey 
offer an excellent combination of optically thin and thick tracers and cover a 
broad range in critical densities and excitation temperatures. 
Table \ref{lines-malt90} summarises the lines included within the survey and their properties (excitation energy and critical density). For simplicity 
throughout the paper we will refer to each transition only by their molecular 
name, i.e., \hcop~instead of \hcop(1-0). 

For each MALT90 clump we extract the line parameters determined via Gaussian profile fitting to the spectra reported in the MALT90 catalog (Rathborne et al, in prep) (velocity, $v_{LSR}$; line width, $\Delta v$; and peak antenna temperature, T$^*_A$). Table 3 shows an example of some of the MALT90 data for the clumps embedded in filament A, the rest of the MALT90 data is available as an online table.

\section{Properties of the embedded clumps}

 \begin{figure*}
 \begin{center}
 \includegraphics[trim = 15mm 105mm 30mm 115mm, clip,width=.99\textwidth]{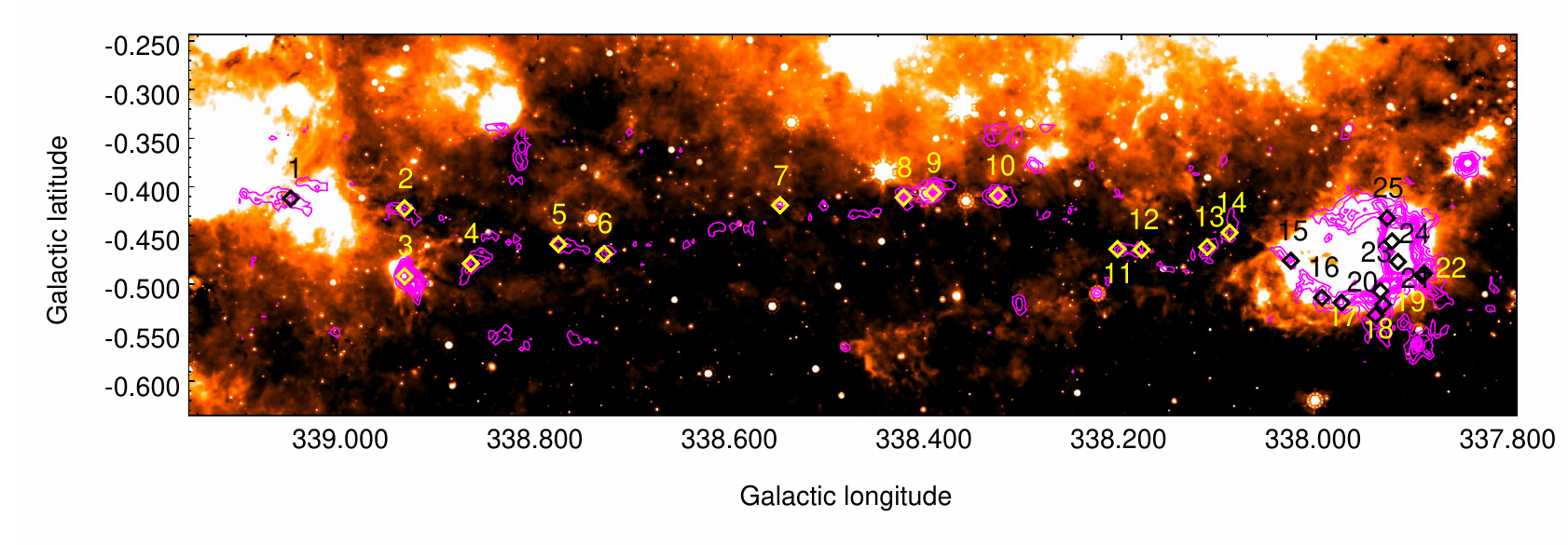}
 \caption{24 \mum~\textit{Spitzer}/MIPS image toward filament A: AGAL337.922-0.456, `Nessie' \citep{Jackson-2010}. The 870 \mum~ATLASGAL emission is overlaid as contours (2 to 10 $\sigma$). Black and yellow diamonds indicate the position of the clumps, identified from the ATLASGAL compact catalog \citep{Contreras-2013a}. The numbers correspond to the ID given to the clumps. For clarity, we have removed the letter in each clump. While most of the clumps have little infrared emission, some show star formation indicators such as compact 8 and 24 \mum~emission.}
 \label{fig:filaments1}
 \end{center}

 \end{figure*}

 \begin{figure*}
 \begin{center}
 \includegraphics[trim = 16mm 100mm 30mm 110mm, clip,width=.99\textwidth]{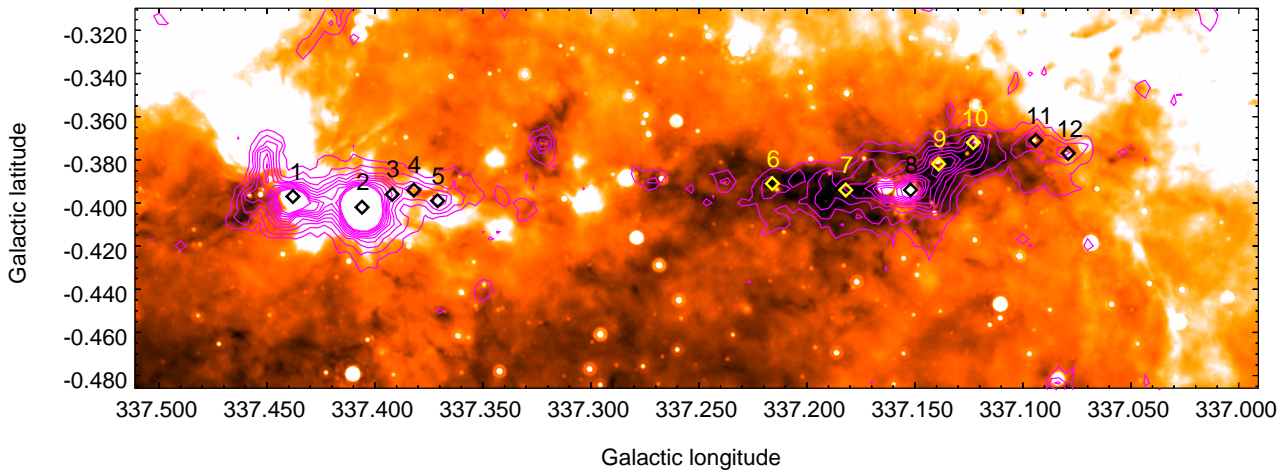}
 \caption{24 \mum~\textit{Spitzer}/MIPS image toward filament B: AGAL337.406-0.402. Contours and symbols as in Figure 1. This filament contains both infrared-dark and infrared-bright clumps.}
 \label{fig:filaments2}
 \end{center}

 \end{figure*}
 \begin{figure*}
 \begin{center}
  \includegraphics[trim = 35mm 100mm 50mm 90mm, clip,width=.8\textwidth]{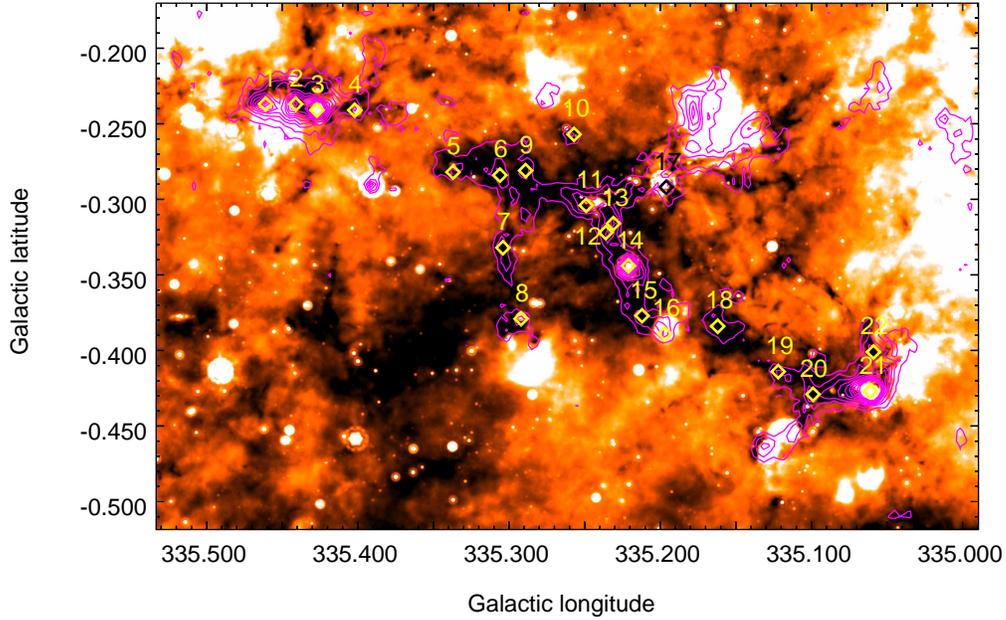}
 \caption{24 \mum~\textit{Spitzer}/MIPS image toward filament C: AGAL335.406-0.402. Contours and symbols as in Figure 1. This filament represents a good example of the hub-tail structure described in \citet{Myers-2011}, showing a main body that run from clump 1 to clump 21 and several ramifications. This filament is also in a young evolutionary stage with many clumps appearing dark in GLIMPSE/MIPSGAL.}
 \label{fig:filaments3}
 \end{center}

 \end{figure*}
 \begin{figure*}
 \begin{center}
  \includegraphics[trim = 21mm 110mm 26mm 105mm, clip,width=.99\textwidth]{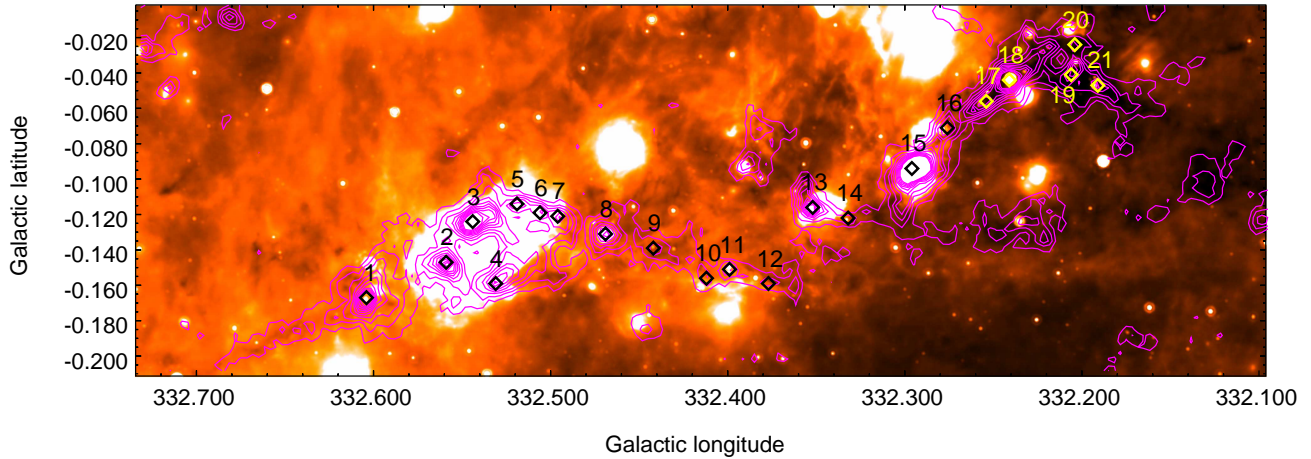}
 \caption{24 \mum~\textit{Spitzer}/MIPS image toward filament D: AGAL332.294-0.094. Contours and symbols as in Figure 1.  This filament is an example of a more evolved structure with several clumps associated with bright infrared emission. Here, we can see that around the HII region clumps (bright region in this image) the clumps are closer, while the rest of the filament the clumps remain more evenly spaced.}
 \label{fig:filaments4}
 \end{center}

 \end{figure*}
 \begin{figure*}
 \begin{center}
  \includegraphics[trim = 20mm 100mm 35mm 90mm, clip,width=.99\textwidth]{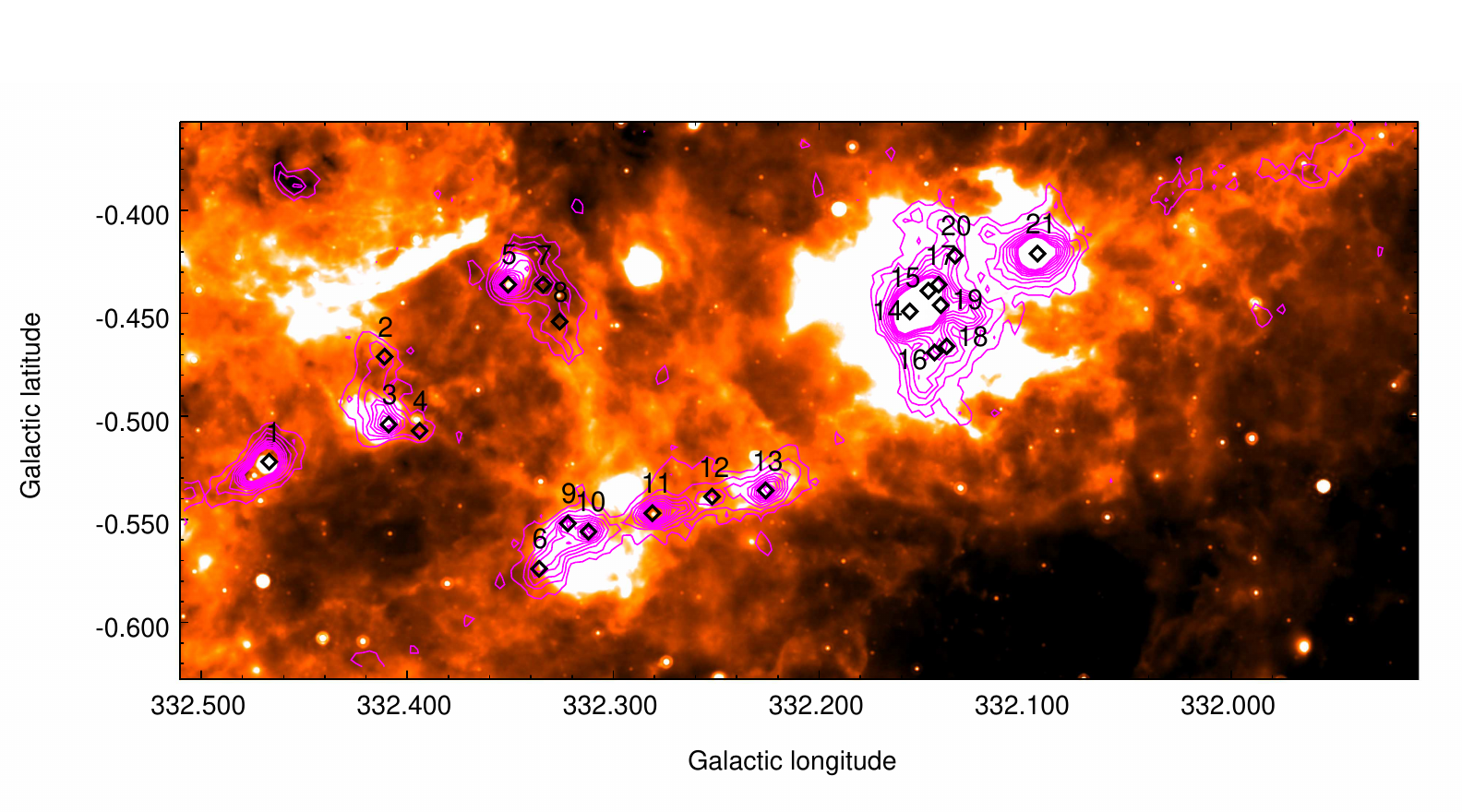}
 \caption[]{24 \mum~\textit{Spitzer}/MIPS image toward filament E: AGAL332.094-0.421. Contours and symbols as in Figure 1. This image shows a small filament composed by 6 clumps (clumps 8-13) associated with bright infrared emission. This is an example of a more evolved filament with all of its clumps showing signs of active star formation.}
 \label{fig:filaments5}
 \end{center}
 \end{figure*}

\subsection{Identification of the clumps}

The 870~\mum~continuum emission was used to pinpoint the location of the clumps along each filament. 
Since this emission traces well both the IR bright and dark regions, it provides 
an homogenous method to determine the position of the clumps. 
The clumps were identified by searching the ATLASGAL 
compact source catalogue for all the clumps located within the more extended 870 \mum~continuum emission that defined each filament \citepalias[see][for further details]{Contreras-2013b}. We found a total of 101 clumps within these five filaments. 
Table \ref{summ-clumps} lists for each clump its abbreviated ID and its full ATLASGAL denomination.

\subsection{Clump sizes, masses, and volume densities}\label{sec:starinfil}

The size, mass, and volume density of the clumps were 
derived from the observed molecular line and dust continuum emission.
The size of a clump was defined as twice the effective radius listed in the ATLASGAL compact source 
catalogue \citep{Contreras-2013a}. Using the kinematic distance to each filament 
(from \citetalias{Contreras-2013b}, listed in Table \ref{summ-frag1}), we obtained the physical size 
for each clump. The sizes range between 0.26 and 1.6 pc, with an average of $\sim$0.5 pc (see Table 4).
The masses of the clumps, derived from their dust continuum emission, were determined using the relation 
\citep{hildebrand-1983}, 
\begin{equation}
M_{dust}=\frac{S_{870} D^2 R_{gd}}{\kappa_{870} B_{870}(T_d)},
\label{dust-mass}
\end{equation}
\noindent where $S_{870}$ is the flux density at 870 \mum, obtained from 
the ATLASGAL compact source catalog, $D$ is the kinematic distance to the source, $R_{gd}$ is the gas to dust mass ratio assumed as 100, 
$\kappa_{870}$ is the dust absorption coefficient, $B_{870}$ is the Plank function at 870 \mum~and $T_d$ is the dust temperature. 
We assumed a dust temperature based on the observed degrees of IR emission
(see section 4.1) and assumed $\kappa_{870}=$1.85 cm$^{2}$ gr$^{-1}$, interpolated from \citet{ossenkopf-1994}. 
With these assumptions we calculated the clump masses and found that they range from $\sim$20 to $~2500$ M$_\odot$ 
(see Table \ref{summ-clumps}). Assuming the clumps have roughly spherical shapes, we compute the H$_2$ volume density via 
$n(H_2)=3M_{dust}/4\pi r^3$, finding values ranging from 
10$^3$ to 10$^5$ cm$^{-3}$ (see Table \ref{summ-clumps}).
 
The virial mass was calculated using the size of the clumps and the line width derived from the 
\nhp~(1-0) emission. The optical depth of the \nhp(1-0)~emission is usually $\lesssim$1, thus, this molecular line is suitable for determining the virial parameter \citep[e.g.][]{Sanhueza-2012}. Optically thick lines would overestimate the virial mass. We found a total of 71 clumps for which there are MALT90 detections of \nhp. For these clumps, we use the line width obtained from the hyperfine structure fitting to compute their virial mass.  The virial mass of a clump of radius $r$ and 
line-width $\Delta {\rm v}$ is given by,
\begin{equation}
\label{mvirial-clumps}
M_{vir}=\frac{5r\Delta {\rm v}^2}{8 ln(2) a_1 a_2 G}\sim 209 \frac{1}{a_1a_2}\left(\frac{\Delta {\rm v}}{{\rm km~s}^{-1}}\right)^2\left(\frac{r}{pc}\right){M_\odot},
\end{equation}
\noindent where $a_1$ is the correction for a power-law distribution given by $a_1=\frac{1-p/3}{1-2p/5}, p<2.5$, and $a_2$ is the correction for a non spherical shape \citep{bertoldi-1992}. To calculate $a_1$ we use a power law density distribution of $p=1.8$. We assume that the clumps in our sample have spherical shapes, therefore we used $a_2=1$ for all clumps. With these assumptions, we found that the clumps have virial masses ranging from $\sim$50 to $\sim$1700 M$_\odot$ (see Table \ref{summ-clumps}).

\section{Evidence for star formation within the clumps}

\subsection{Infrared signatures of star formation}

To characterize the star formation activity within each clump, we used 
\textit{Spitzer} 24 \mum~MIPSGAL and 3.6 - 8 \mum~GLIMPSE surveys.

Using these IR images we classified the clumps into four broad categories: 
`pre-stellar' clumps, appearing dark at 3.6-8 \mum, and 24 \mum; `proto-stellar' 
clumps, having a 24 \mum~point like emission that traces the dust 
heated by embedded proto-stars, or enhancement in the emission at 4.5 \mum, indicative of the presence of shocked gas \textbf{(often referred too in the literature as extended green objects, or `green fuzzies', for their green appearance in a composite image where green is used to represent 4.5 $\mu$m emission}); `HII region' clumps, 
associated with bright, compact 8 and 24 \mum~emission; and `PDRs clumps' showing extended 
and often diffuse 8 and 24 \mum~emission. These categories broadly describe the evolution of a clump since the IR emission associated within a clump will increase as their embedded stars form and heat their immediate surroundings. \citep[][]{Chambers-2009,jackson-2013}. Using this scheme we categorized 31 of the clumps as `pre-stellar' (30\%), 30 as `proto-stellar' (30\%), 11 as `HII region' clumps (11\%), and 29 as PDRs (29\%).

For each clump we assumed a dust temperature equal to the average temperature of it classification obtained by Guzman et al. (in press.) for a large sample of clumps located in the Galactic plane. Guzman et al. derived dust temperatures for all MALT90 clumps, by fitting their dust continuum emission using data from the Hi-GAL and ATLASGAL surveys. The average dust temperatures are 17 K for `pre-stellar' clumps, 19 K for `proto-stellar' clumps, 24 K for clumps associated with `HII regions', and 28 K for PDRs. Since, not all the clumps detected in the filaments have MALT90 observations, in this paper we use the average dust temperature derived for the relevant category in calculations for each of the individual clumps. Table \ref{summ-clumps} summarizes the properties of the IR emission detected towards each clump as well as 
their IR-based classification and assumed dust temperature.

We found that the amount of IR emission and the number of clumps assigned to each category differed between the filaments, suggesting that the filament span a range of evolutionary stages. In filaments A and C, most of the clumps are IR dark. In filament B, some of the clumps are dark and others are IR bright. In filaments D and E, most of the clumps are associated with `HII regions' and PDRs.

\subsection{Molecular line signatures of star formation} 

\subsubsection{Detection rates and evolutionary stage}

 \begin{figure*}
 \begin{center}
 \includegraphics[trim = 0mm 27mm 0mm 30mm, clip,width=.9\textwidth]{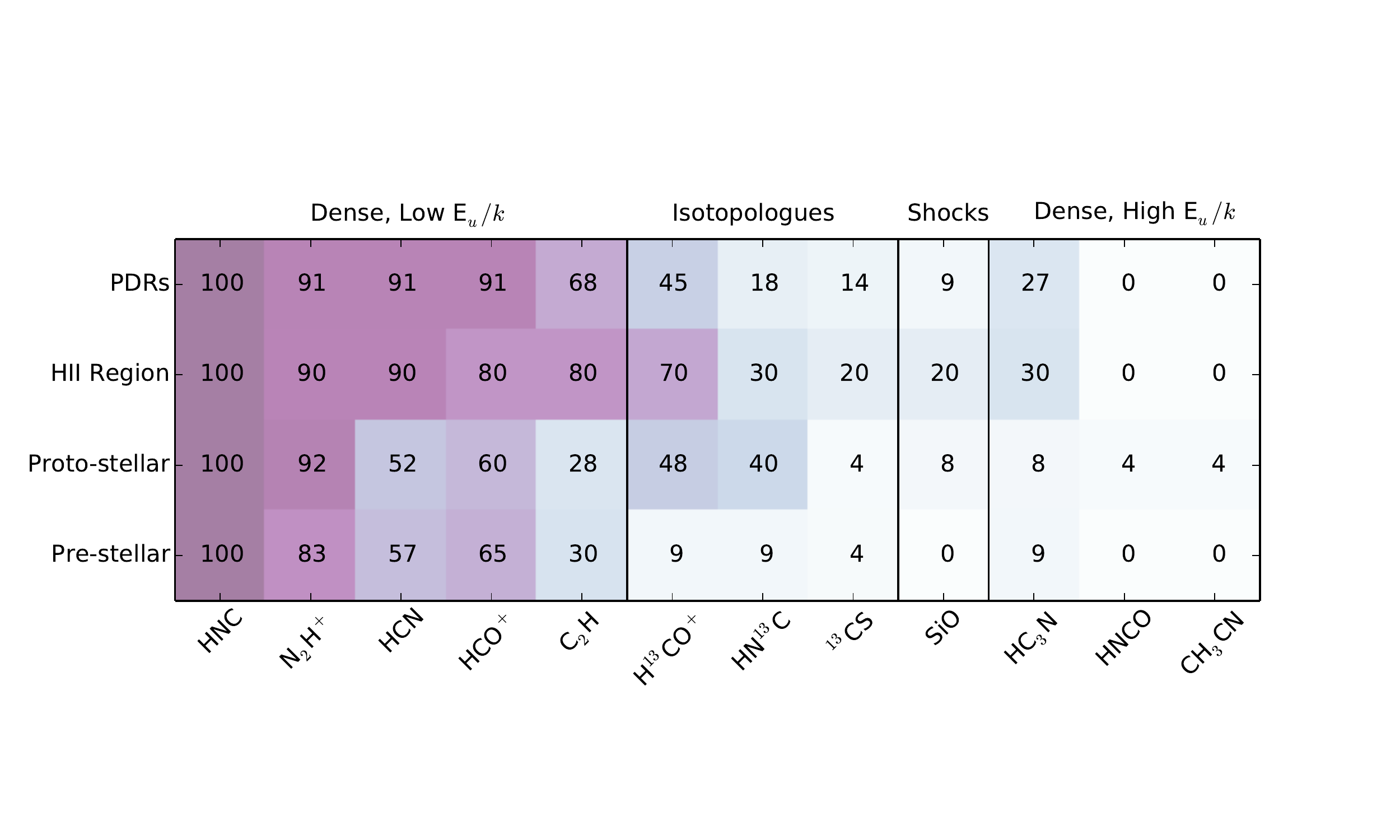}
 \caption{Detection rate of the MALT90 clumps. Numbers indicate the percentage of clumps for which each molecule was detected.}
 \label{fig:detecrates}
 \end{center}
\end{figure*}

An embedded proto-star influences the gas chemistry 
of its immediate surrounding as it forms and evolves.  In the initial stages, a clump's interior will have low temperatures and high density, where molecules are stuck on the icy 
dust grains. As a proto-star evolves and starts heating its surrounding environment, 
molecules will be released into the gas phase and their emission will be 
detected. 

MALT90 observations were used to determine the molecular line detection rates towards the clumps and
their relationship with their star formation activity. Figure \ref{fig:detecrates} summarizes the detection rates for each molecule for the MALT90 clumps broken into their IR-based categories. We found that the number of molecular species detected increases from the clumps classified as `pre-stellar' to the clumps associated with `HII Regions' and PDRs, which reflects the increased complexity in the chemistry as the embedded proto-stars within these clumps form and evolve.
 
Of the four main dense gas tracers, emission was detected in \hnc~(above 3$\sigma$) towards all 80 MALT90 clumps, in \nhp~towards 89\%, in HCO$^+$ towards 73\% and in HCN towards 69\% of the MALT90 clumps. These molecules are typical tracers of high density gas and are detected over a broad 
range of gas temperatures \textbf{and} evolutionary stages \textbf{\citep[e.g.,][]{Sanhueza-2012, Hoq-2013}}. We find that, in general, the morphology of the \nhp~emission traces very well the 
clumps, while the \hnc~emission is more extended tracing well the overall morphology 
of the filaments.

The reactive radical ethynyl (C$_2$H) has been detected in quite different environments such as those towards low-mass cores \citep{millar-freeman-1984}, PDRs \citep{Jansen-1995}, high-mass star forming regions \citep{Beuther-2008} and infrared dark clouds \citep{Sanhueza-2012, Sanhueza-2013}. For the MALT90 clumps, \ch~was detected toward 30\% of the `pre-stellar' clumps, 28\% of the `proto-stellar' clumps, 80\% of the `HII region' clumps and in 68\% of the PDRs. We find that the morphology of the C$_2$H emission matched well the morphology of the HNC and N$_2$H$^+$ emission towards the `pre-stellar' and `proto-stellar' clumps. In contrast, however, the C$_2$H emission associated with the PDR clumps appears to more closely follow the extended, diffuse background emission.  

Emission from the isotopologues \hntc\,, \htcop\ and \tcs~was detected predominantly towards the `proto-stellar' clumps and `HII region' clumps, \htcop\ being the most frequently detected isotopologue. SiO was detected  toward six clumps: two `proto-stellar', two `HII region' clumps, and two PDRs. SiO is usually associated with outflows, because the Si atoms are released from dust into the gas phase by shocks produced by outflows from newly formed stars. Since the presence of enhanced 4.5 \mum~emission (relative to 3 and 8 \mum) may also arise from shocked gas, one might expect a tight correlation between those clumps with extended 4.5 \mum~emission and the presence of SiO. We found that about 10\% of the 4.5 \mum~bright clumps have detected SiO emission. This low detection rate of SiO towards these clumps may reflect the MALT90 sensitivity limit, the differences in the angular resolution of the two data sets (the MALT90 data will be significantly beam diluted), or it may indicate that the proto-stars in these clumps are in an early evolutionary stage, and they
have not yet evolved to produce outflows that sufficiently excite the SiO.

Less frequently detected was the emission from ``hot core chemistry" molecules such as \hctn~(16\%), CH$_3$CN (4\%) and \hnco (4\%). \hctn~was detected in 9\% of the `pre-stellar' clumps, 8\% of the `proto-stellar' clumps, 30\% of the `HII region' clumps and 27\% of the PDRs. As expected, we find that the detection rate for these more complex molecules increases towards the more evolved clumps. CH$_3$CN and \hnco~were only detected towards four clumps, all of these clumps were classified as `proto-stellar'. No emission was detected in the HNCO (J$_{K_a,K_b}$=4$_{1,3}$-3$_{1,2}$), HC$^{13}$CCN, $^{13}$C$^{34}$S and H${41\alpha}$ lines toward any of the 80 MALT90 clumps. 

\section{Discussion}

\subsection{The fragmentation of filaments via the `sausage' instability}

\begin{table*}
\begin{minipage}{0.99\textwidth}
\caption{Summary of the relevant properties of the filaments used in this work. For more details refer to \citetalias{Contreras-2013b} on the values of the distance (D), Radius of the filament (r), line width ($\Delta v$), central density ($\rho_c$), inner flat region radius ($R_{flat}$) and the index of the intensity profile that relates to the magnetic field support ($p$). NC is the number of clumps detected in each filament. $\lambda_{obs}$ is the observed spacing between the clumps in each filament, with the error give by the standard deviation between the separation between consecutive clumps and the average observed separation. $H_{eff,0}$ and $\lambda_{s,0}$ correspond to the non magnetic solution, i.e. $\gamma=0$. B$_c$ is the value of the magnetic field that produce a $\lambda_{s}$ = $\lambda_{obs}$.}
\label{summ-frag1}
  \begin{center}
    \begin{tabular}{l|cccccccccccc}
      \hline
      \hline
      Filament &D&\multicolumn{1}{|c}{r} &NC& \multicolumn{1}{|c}{$\Delta v$}&\multicolumn{1}{|c}{$\rho_c$}&\multicolumn{1}{|c}{R$_{flat}$}&\multicolumn{1}{|c}{$p$} &\multicolumn{1}{|c}{H$_{eff,0}$}&\multicolumn{1}{|c}{$\lambda_{obs}$}&\multicolumn{1}{|c}{$\lambda_{s,0}$}&\multicolumn{1}{|c}{$B_{c}$} \\ 
       &&&&&x10$^{-19}$&&&\multicolumn{1}{|c}{}&&& \\ 
       & \multicolumn{1}{|c}{(kpc)} & \multicolumn{1}{|c}{(pc)} & &\multicolumn{1}{|c}{(\kms)} & \multicolumn{1}{|c}{(gr cm$^{-3}$)}& \multicolumn{1}{|c}{(pc)} && \multicolumn{1}{|c}{(pc)}& \multicolumn{1}{|c}{(pc)}  & \multicolumn{1}{|c}{(pc)}&\multicolumn{1}{|c}{($\mu G$)}\\
      \hline
      \hline
      A& 3.0&0.44 &25& $^a$2.8 & 0.74&0.25&3.0&0.155&3.2 $\pm$ 2.4&3.4&110\\
      B& 3.2&0.45 &12& 2.4 & 1.7&0.30&2.0&0.091&1.9 $\pm$ 2.1&2.0&128\\
      B1&3.2& 0.45 &5& 2.4 & 1.7&0.30&2.0&0.091&1.3 $\pm$ 0.4&2.0&537\\
      B2&3.2& 0.45 &7& 2.4 & 1.7&0.30&2.0&0.091&1.0 $\pm$ 0.5&2.0&1031\\
      C&3.6& 0.56 &22& 2.7 & 1.2&0.64&3.1&0.116&2.4 $\pm$ 1.7&2.6&133\\
      D&3.3& 0.52 &21& 2.1 & 1.3&0.58&2.0&0.086&1.7 $\pm$ 0.7&1.9&151\\
      E&3.7& 0.23 &21& 3.3 & 3.0&0.50&2.6&0.091&1.6 $\pm$ 0.5&2.0&580\\
      \hline
      \hline
    \end{tabular}
    \medskip
  \end{center}
$^a$ We used here the mean value of the \nhp~emission, since the \tco~observations toward this filament don't cover their full extent (see Paper I for more details).\\
\end{minipage}
   \end{table*}

The process leading to the formation of clumps in filaments is yet to be understood, 
but theoretical work shows that filaments can fragment due to gravitational and MHD driven instabilities \citep{fiege-pudritz-2000, chandrasekhar-1953, nagasawa-1987, 
inutsuka-1992, nakamura-1993, tomisaka-1995}. Because the selected filaments are consistent with being confined by a helicoidal, toroidal-dominated magnetic field \citepalias[see][]{Contreras-2013b}, we consider the fragmentation of these filaments in the framework of the MHD instabilities within filaments that are also confined by such magnetic fields. In particular, our goal is to determine whether or not the `sausage' MHD instability is able to accurately describe the observed spacing between the clumps within each of the filaments. 

The selected filaments can be represented by an isothermal finite cylinder with an helicoidal magnetic field. In this case,
\citet{nakamura-1993} predicted a preferred spacing given by,
\begin{equation}
\lambda_{s}= 8.73 H [(1+\gamma)^{1/3}-0.6]^{-1} ~~,
\end{equation}
\noindent where $\gamma= B_c^2/(8\pi\rho_cc_s^2)$, $c_s$ is the sound speed, 
$B_c$ is the magnetic field strength, $\rho_c$ is the central density of the 
filament, and H is the scale of height, given by, 
\begin{equation}
	\label{eq:scale-height}
  H=c_s(4\pi G\rho_c)^{-1/2}[1+\gamma/4]^{1/2},
\end{equation}
\noindent where G is the gravitational 
constant. Thus, when magnetic fields are considered, the expected spacing between 
the clumps is predicted to be smaller than the pure non-magnetic case.

Because the observed line-widths of the molecular tracers detected towards these filaments show that the turbulent pressure dominates over the thermal 
pressure, the sound speed $c_s$ in Equation 4 should be replaced by 
the velocity dispersion $\sigma$ \citep{fiege-pudritz-2000} of the gas in the 
filament, hence:
\begin{equation}
	\label{eq:scale-height}
  H_{eff}=\sigma(4\pi G\rho_c)^{-1/2}[1+\gamma/4]^{1/2},
\end{equation}
\noindent where $\gamma= B_c^2/(8\pi\rho_c\sigma^2)$. If $\gamma=0$, thus $\lambda_{s,0}=21.8 H_{eff,0}$ with $H_{eff,0}=\sigma(4\pi G\rho_c)^{-1/2}$.

The central densities of the filaments were estimated from the fit to the 870 \mum~dust continuum emission, as 
described in \citetalias{Contreras-2013b} assuming an intensity profile given by \citep{Arzoumanian},
\begin{equation}
  \label{eq:lineal-prof-2}
  I(r)=A_p\frac{\rho_cR_{flat}}{[1+(r/R_{flat})^2]^{\frac{p-1}{2}}}
\frac{k_\lambda\ B_\lambda(T_{dust})}{d^2},
\end{equation}

\noindent where I(r) is the dust continuum emission intensity profile; p is an index that gives an 
indirect measurement of the magnetic field support over the filament (in the non-magnetic case p = 4;  \citet{ostriker-64}; and $p < 3$ if the filament 
is magnetically supported, \citet{fiege-pudritz-2000}); $A_p$ is a constant given by 
\mbox{$A_p=\int_\infty^\infty du/(1+u^2)^{p/2}$} for p$>$1, which is equal to $\pi/2$ for the non-magnetic case;
$R_{flat}$ is the radius of an inner flat region, which in the non-magnetic case corresponds to the thermal Jeans length at
the centre of the filament; $k_\lambda$ is the dust opacity; 
$B_\lambda$ is the Planck function; $T_{dust}$ is the average dust temperature 
of the filament, assumed to be 20 K; and $d$ is the distance to the filament. The values 
of $\rho_c$, R$_{flat}$, $p$, and $H_{eff,0}$ for these filaments are summarized in Table \ref{summ-frag1}.
 
Using the ATLASGAL maps we measured the separation 
between a clump and its nearest neighbour clump. Because the velocity gradient in the filaments is small, we assume that the filaments are perpendicular to the line of sight when calculating the observed spacing. 

The average value of the observed spacings, $\lambda_{obs}$, between the clumps range from 1.0 to 3.2 (see Table \ref{summ-frag1}). To measure how periodic is the location of the clump within the filament, the error in the observed spacing was calculated by taking the standard deviation between the separation of consecutive clumps and the average observed spacing. Filament B has weak dust 870 \mum~dust continuum emission towards its centre, creating two sections within this filament. If we treat each section independently (B1 and B2, in Table \ref{summ-frag1}), then the clump observed spacing in each section is more periodic (the associated error in each section is smaller). 

The  theoretical spacing predicted for each filament assuming $\gamma=0$, $\lambda_{s,0}$, range between 1.9 and 3.4 pc (see Table 
\ref{summ-frag1}). In general $\lambda_{s,0}$ is slightly higher than the observed spacings in all the filaments. This discrepancy could be 
attributed to the presence of magnetic fields.
We find that to explain the observed spacing, the `sausage' instability theory requires magnetic fields ranging between 
$\sim$110 and 580 $\mu$G.

Regardless of their complex morphology or evolutionary stage, we found that for all filaments the predicted spacing given by the `sausage' instability 
theory and the observed spacing matches well for the case of a helical magnetic field with strength of hundreds $\mu$G. This suggests that this mechanism could work not only for `ideal' filaments such as `Nessie' but also for filaments 
that have more complex morphologies. The average spacing of the more evolved 
filaments is also similar to the predicted initial spacing given by this theory. This would imply then, that if the clumps are formed via this mechanism, as they evolve into HII regions and PDRs and disrupt 
their environment, the average spacing along the whole 
filament is not significantly affected over time. This means that we can trace the fragmentation of the filaments, even if they are evolved, by observing their sub-millimetre continuum emission.

\subsection{Star formation within the filaments}

From the observed IR emission detected towards the clumps, 
we were able to estimate the overall star formation activity within the filaments. 
Filaments A, B and C have a higher number of clumps in the `pre-stellar' phase, 
while filaments E and D contain a relatively higher number of clumps in more evolved evolutionary stages 
(i.e., `HII region' clumps and PDRs). In this section we discuss the virial state of the clumps embedded and their potential to form high-mass stars.

\subsubsection{Are the clumps gravitationally bound?}

 \begin{figure*}
 \begin{center}
 \includegraphics[trim = 0mm 0mm 0mm 0mm, clip,width=.6\textwidth]{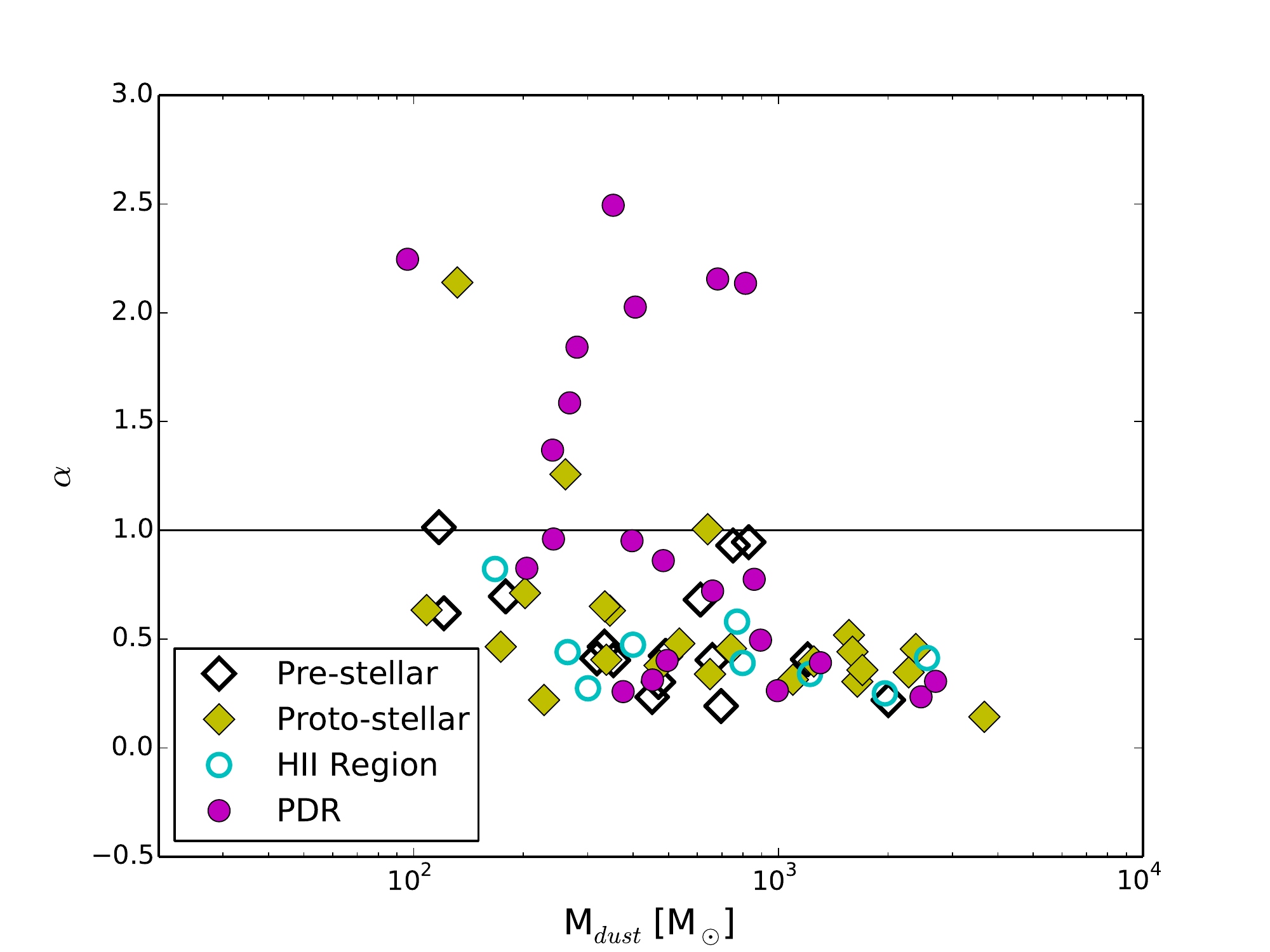}
 \caption{Virial parameter for the clumps in each evolutionary stage. Most of the `pre-stellar' and `proto-stellar' clumps have values of $\alpha<1$, suggesting that they are gravitationally bound and likely to form stars. PDRs have a wider range in values consistent with this late evolutionary stage where clumps are expected to expand.}
 \label{fig:alpha}
 \end{center}
\end{figure*}

To determine whether the clumps are gravitationally bound, we calculated the virial parameter $\alpha$ (ratio between 
the virial and dust mass) for those clumps with molecular line 
information. The parameter $\alpha$ is commonly used to determine if a clump 
is stable against collapse. If $\alpha>1$ the clump contains enough kinetic 
energy to expand unless it is confined by an external mechanism. A value 
of $\alpha<1$ indicates that the kinetic energy is not enough to 
support the clump against gravitational collapse \citep{Kauffmann-2013}. 
 
We find that 83\% of MALT90 clumps have values of $\alpha\leq1$ and are unstable to gravitational collapse, if magnetic fields are ignored.
Figure \ref{fig:alpha} shows the values of the virial parameter for all the clumps colour coded by their IR-based categories. Most of the `pre-stellar' and `proto-stellar' clumps have virial parameter with values $<1$, suggesting that there are gravitationally bound and likely to collapse. This is supported with the fact that `proto-stellar' clumps are currently forming stars, and thus are likely collapsing. Only one `pre-stellar' clump and two `proto-stellar' clumps have values of $\alpha \geq 1$, thus, they might correspond 
to transient clumps. We find that all the `HII region' clumps also have `virial parameter' lower than 1, suggesting that their recently formed proto-star has not yet significantly affected the clump dynamics. The PDRs have virial parameters that have a wider range in values, which is also expected for these late evolutionary stages where the expansion of the clumps is expected due to the UV radiation and stellar winds from the newly formed high-mass stars.

\subsubsection{Will the clumps form high-mass stars?}

 \begin{figure*}
 \begin{center}
 \includegraphics[trim = 0mm 0mm 0mm 0mm, clip,width=.7\textwidth]{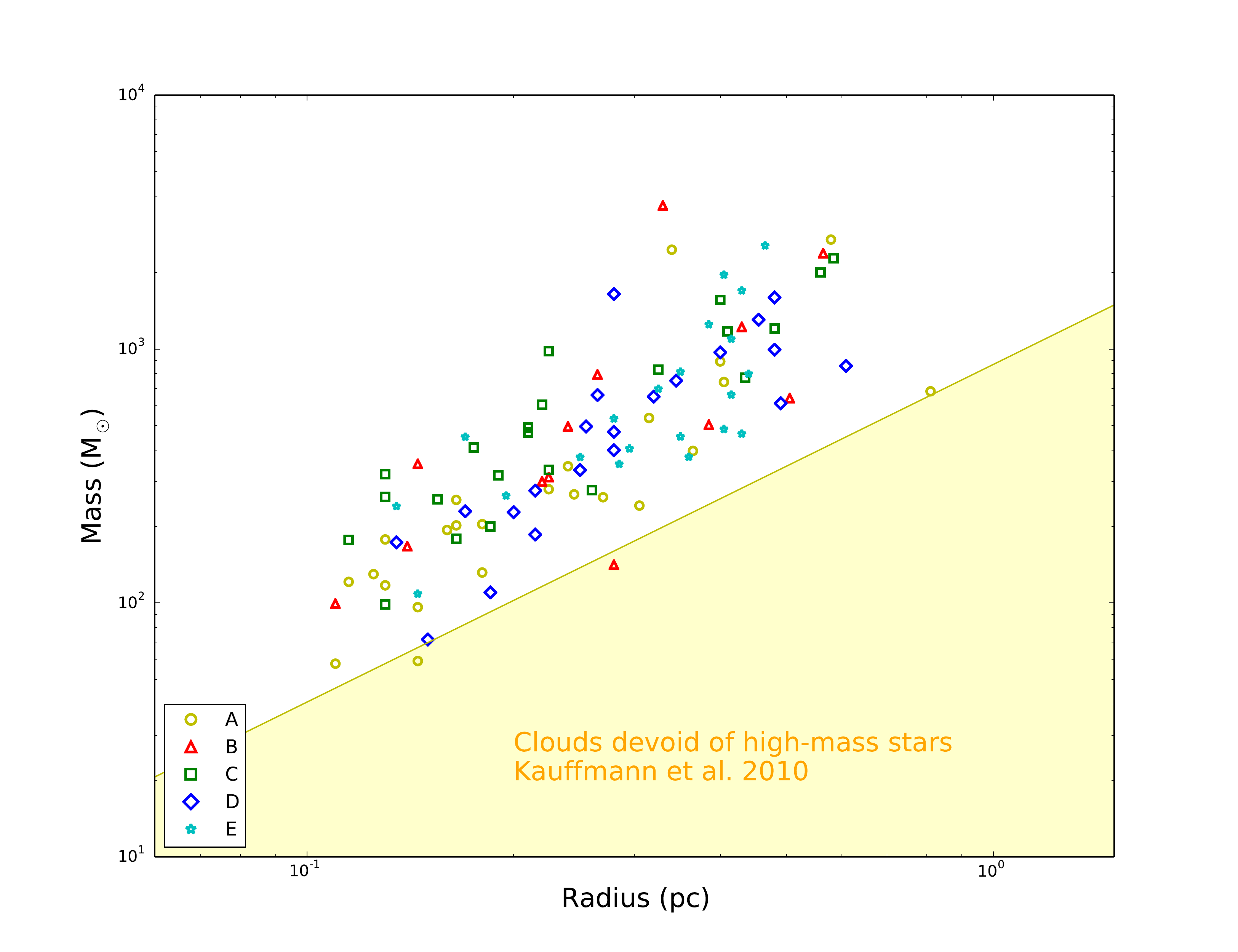}
 \caption{Mass versus radius for the clumps embedded in the filaments. The clumps for different filaments are shown in different colours and symbols. The legend shows the colours and symbols used for each filament. The yellow region shows the parameter space for clumps devoid of high-mass stars. All but 2 of our clumps lie above the region of clumps devoid of high-mass stars, suggesting that all the filaments will harbour high-mass stars.}
 \label{fig:massradius}
 \end{center}
\end{figure*}

The total mass of the clumps in each filament, are $1.3\times 10^4$, $1.1\times10^4$, $1.5\times 10^4$, $1.3\times 10^4$ and $6.2\times10^3$ M$\odot$ for filaments A, B, C, D and E respectively. To determine the potential of these clumps to form high-mass stars ($>$8 M$\odot$), we conservatively assumed that 30\% of the mass of the clumps will form cores, and that only 10\% of the mass in the cores will eventually be converted into stars. Given a typical initial mass function \citep[IMF; e.g.][]{Salpeter-1955}, we determined that all the filaments will have enough material to form more than one star with a mass $>8$ M$\odot$. 

While it appears that the clumps have sufficient mass to form at least one high-mass star, we must also consider their sizes and volume densities. Figure \ref{fig:massradius} shows the mass versus radius plot for all the 101 clumps embedded in the five filaments. We compare the mass and sizes of our clumps with the limits established by \citet{Kauffmann-2010b} for clumps devoid of high-mass star formation, which follow the relationship $M(r)\leq 870 M_\odot (r/pc)^{1.33}$. We find that 98\% of the clumps lie above the threshold for clump devoid of high-mass stars. This provide additional support to the idea that these clumps  will harbour high-mass stars.

\section{Summary}
Using data from the ATLASGAL, \textit{Spitzer}, and MALT90 surveys we characterized the physical properties, chemistry, kinematics, and 
star formation activity of 101 clumps embedded in five filamentary molecular 
clouds. We find that each filament harbour several clumps, covering a wide range of masses 
from hundred to thousands solar masses.

We studied the emission of dense gas tracers towards 80 clumps using MALT90 data. We find that the number of lines detected towards the clumps increases with the amount of associated IR emission and, hence, active star formation. Emission in the HNC line was detected towards all 80 clumps. N$_2$H$^+$, HCO$^+$, HCN and C$_2$H were detected toward 89\%, 73\%, 69\% and 46\% of the clumps, respectively. For clumps with classified as `pre-stellar' and `proto-stellar', the 
morphology of the C$_2$H emission is similar to that of the dust 
emission and other molecules. For clumps 
classified as PDR, C$_2$H traced well the morphology observed in the PDRs. The isotopologues \hntc, \htcop, and \tcs~were detected more frequently towards `proto-stellar' clumps, `HII region' clumps and PDRs than `pre-stellar' clumps. SiO emission was detected towards $\sim10$\% of the MALT90 clumps and less frequently detected was emission from the ``hot core chemistry" molecules. 

Assuming that the filaments are isothermal finite cylinders supported by a helicoidal magnetic field, we find that the observed separation of the clumps can be explained via the `sausage' instability theory for magnetic field in the range from $\sim$110 to 580 $\mu$G. Moreover, we found that regardless of the morphology and evolutionary stage 
exhibited by the filaments, by following the dust continuum 
emission we can still trace the underlying structure within the filament. 

The derived virial parameter for the clumps reveals that they are typically gravitationally 
bound. Given the clump masses and assuming a typical IMF, we found that every filament host a clump that is 
likely to form at least one high-mass star. We also analysed the mass versus radius relationship of the clumps, finding that most of the clumps are likely to form high-mass stars. This result further supports the idea that filaments are the birthplaces of high-mass stars and their associated clusters.

\section*{Acknowledgments}
We thank the referee Michael Burton for his valuable comments that significantly improved this paper. YC and GG gratefully acknowledge support from CONICYT through projects FONDAP No. 15010003 and BASAL PFB-06.

\bibliography{bibliografia}
\clearpage
\newpage

\begin{table*}
  \caption[List of the molecular lines observed]{List of the molecular transitions observed as part of MALT90.}
  \label{lines-malt90}
\begin{center}
\begin{tabular}{lccccll}
  \hline
  \hline
  Molecule &Transition&	Frequency & $n_{crit}$& E$_u/k$ &	Tracer\\
  && (MHz)     &(cm$^{-3}$)&(K)& &\\
  \hline
  \hline
  
  H$^{13}$CO$^+$&J=1-0	& 86754.330	&2$\times 10^5$&4.16& Column density            \\
  HN$^{13}$C&J=1-0	& 87090.735	&3$\times 10^5$&4.18& Column density            \\ 
C$_2$H&N=1-0,J=3/2-1/2, F=2-1	& 87316.925	&2$\times 10^5$&4.19&  Density, Photodissociation tracer  \\
HCN	& J=1-0        &88631.847	&3$\times 10^6$&4.25& Density                   \\
HCO$^+$	& J=1-0        &89188.526	&2$\times 10^5$&4.28& Density                   \\
HNC	 & J=1-0        & 90663.572	&3$\times 10^5$&4.35&Density; cold chemistry   \\
  N$_2$H$^+$   &J=1-0	& 93173.480	&3$\times 10^5$& 4.47& Density, cold chemistry\\
  SiO&J=2-1	& 86847.010	&2$\times 10^6$&6.25& Shock/outflow tracer      \\

  $^{13}$CS  &J=2-1	& 92494.303	&3$\times 10^5$&  6.66& Column density            \\
  $^{13}$C$^{34}$S& J=2-1	& 90926.036	&4$\times 10^5$&7.05& Column density            \\
  HNCO &J$_{K_a,K_b}$=4$_{0,4}$-3$_{0,3}$&	88239.027&1$\times 10^6$&10.55&	Hot core, shocks            \\
  
  CH$_3$CN  &J$_K$=5$_1$-4$_1$& 91985.316	&4$\times 10^5$&20.39& Hot core                  \\
  HC$_3$N   &J= 10-9	& 91199.796	&5$\times 10^5$&24.01&Hot core                  \\

  HC$^{13}$CCN&J=10-9, F=9-8& 90593.059	&5$\times 10^5$&24.37& Hot core                  \\

  HNCO &J$_{K_a,K_b}$=4$_{1,3}$-3$_{1,2}$&	87925.238&1$\times 10^6$&53.86&	Hot core, shocks        \\

  H         &41$\alpha$   & 92034.475	&&& Ionised gas               \\

\hline
\hline
\end{tabular}
\end{center}

\end{table*}

        \begin{table}
\begin{center}
        \scriptsize

        \caption{Parameter derived from molecular line observations. This table in an online only table that contains the velocity, peak temperature and line width and their errors for all the molecules detected in our sample. Here we show a extract of the table, showing the parameters of the N$_2$H$^+$ emission detected toward the clumps embedded in filament A.}
        \label{res-obs-malt90}

        \begin{tabular}{@{}cccccccccc}

        \hline\hline

        Source & \multicolumn{6}{c}{N$_2$H$^+$} \\ 

        & $v_{\rmn{LSR}}$& $\delta v_{\rmn{LSR}}$&T$_{\rmn{A}}$ &$\delta$T$_{\rmn{A}}$ & $\Delta v$& $\delta\Delta v$\\ 

        & (km s$^{-1}$)& (km s$^{-1}$) & (K)& (K) & (km s$^{-1}$) & (km s$^{-1}$)\\ 

        \hline 

A1  & -35.40 & 0.19  & 0.22 & 0.04  & 2.38 & 0.44  \\
A3  & -36.85 & 0.03  & 1.49 & 0.04  & 2.39 & 0.06  \\
A4  & -36.06 & 0.09  & 0.60 & 0.04  & 2.88 & 0.18  \\
A5  & -39.90 & 0.21  & 0.17 & 0.02  & 3.27 & 0.40  \\
A8  & -37.19 & 0.05  & 0.77 & 0.03  & 2.44 & 0.10  \\
A9  & -37.53 & 0.07  & 0.71 & 0.05  & 2.36 & 0.17  \\
A10 & -38.10 & 0.06  & 0.88 & 0.04  & 2.49 & 0.13  \\
A11 & -37.79 & 0.09  & 0.43 & 0.04  & 2.50 & 0.20  \\
A12 & -38.39 & 0.06  & 0.56 & 0.04  & 2.11 & 0.15  \\
A16 & -39.35 & 0.19  & 0.34 & 0.04  & 3.44 & 0.39  \\
A17 & -39.65 & 0.05  & 0.86 & 0.04  & 2.28 & 0.11  \\
A18 & -35.45 & 0.09  & 0.80 & 0.03  & 3.96 & 0.16  \\
A19 & -37.59 & 0.04  & 1.14 & 0.03  & 3.19 & 0.08  \\
A20 & -38.52 & 0.03  & 1.19 & 0.03  & 2.75 & 0.07  \\
A21 & -39.70 & 0.03  & 1.92 & 0.05  & 2.53 & 0.07  \\
A22 & -39.70 & 0.03  & 1.68 & 0.04  & 2.66 & 0.06  \\
A23 & -38.79 & 0.06  & 0.85 & 0.03  & 3.41 & 0.11  \\
A24 & -40.13 & 0.03  & 2.12 & 0.04  & 3.12 & 0.06  \\
A25 & -38.96 & 0.05  & 1.34 & 0.04  & 3.52 & 0.09  \\

         \hline
        \hline
    
        \end{tabular}
    \end{center}
        \end{table}

\begin{table*}
  \caption{Summary of the properties of the clumps embedded within the filaments. Column 1 and 2 list the IDs given to the clumps and their ATLASGAL name respectively. Column 3 lists the signatures detected in the infrared images. Column 4 lists the IR-based category assigned to each clump (for more details see section 4.1). Column 5 lists IRAS sources that are located within 30" from the clump. Column 6 lists the size of the clumps. Column 7 lists the assumed dust temperatures. Column 8 lists the mass derived from the dust emission. Column 9 lists the virial mass derived from the \nhp~line-width. Column 10 lists the ratio between the virial and dust mass and Column 10 lists the volume density.}\label{summ-clumps}

  \begin{center}
  {\scriptsize
    \begin{tabular}{l|cllccccccr}
      \hline
      \hline
      Clump &\multicolumn{1}{|c}{ATLASGAL} & \multicolumn{1}{|c}{Infrared emission}&\multicolumn{1}{|c}{Classification}&\multicolumn{1}{|c}{IRAS} 
       & \multicolumn{1}{|c}{Size}& \multicolumn{1}{|c}{Dust} & \multicolumn{1}{|c}{M$_{dust}$} & \multicolumn{1}{|c}{M$_{vir}$}&\multicolumn{1}{|c}{$\alpha$}&\multicolumn{1}{|c}{$n(H_2)$}\\
       & \multicolumn{1}{|c}{denomination}&& & &  &\multicolumn{1}{|c}{ Temp.$^b$ }& & &&\multicolumn{1}{|c}{x 10$^4$}\\
       &             && & &\multicolumn{1}{|c}{ (pc)} &\multicolumn{1}{|c}{ (K)} & \multicolumn{1}{|c}{(M$_\odot$)}& \multicolumn{1}{|c}{(M$_\odot$)}&& \multicolumn{1}{|c}{(cm$^{-3}$)}\\
      \hline
      \hline
        
        A1  &AGAL339.054-00.412&24 \mum~emission&Proto-stellar&...&0.33&19							&338		&137&0.41&54.2\\ 
        A2  &AGAL338.937-00.422&4.5 \mum-e, 24 \mum-p$^c$&Proto-stellar&...&0.32&19					&194		&...&... &34.1\\
        A3  &AGAL338.937-00.492&24 \mum-p&Proto-stellar&...&0.81&19						&742		&339&0.46 &8.1\\
        A4  &AGAL338.869-00.479&4.5 \mum-e, 24 \mum-p&Proto-stellar&...&0.54&19					&261		&327&1.25 &9.6\\
        A5  &AGAL338.779-00.459&4.5 \mum-e, 24 \mum-p&Proto-stellar&...&0.36&19					&132		&281&2.14&16.3\\   
        A6  &AGAL338.732-00.469&Dark&Pre-stellar&...&0.25&17										&130		&...&...&47.9\\ 								
        A7 &AGAL338.551-00.419&4.5 \mum-e, 24 \mum-p&Proto-stellar&16396-4631&0.22&19				&58		&...&...	&31.2\\  
        A8 &AGAL338.424-00.411&4.5 \mum-e, 24 \mum-p&Proto-stellar&...&0.33&19					&202		&144&0.71	&32.4	\\
        A9 &AGAL338.394-00.406&4.5 \mum-e, 24 \mum-p&Proto-stellar& 16390-4637&0.63&19			&535		&257&0.48	&12.4\\ 
        A10 &AGAL338.327-00.409&4.5 \mum-e, 24 \mum-p&Proto-stellar&...&0.48&19					&345		&218&0.63&18.0\\	
        A11 &AGAL338.199-00.464&Dark&Pre-stellar&...&0.26&17										&117		&119&1.01&38.5\\				 
        A12 &AGAL338.182-00.464&Dark&Pre-stellar&...&0.23&17										&121		&75&0.62&57.4\\				 
        A13 &AGAL338.112-00.464&Dark&Pre-stellar&...&0.33&17										&254		&...&...&40.9\\				 
        A14 &AGAL338.089-00.447&Dark&Pre-stellar&...&0.26&17										&177		&...&...&58.4\\                 
        A15 &AGAL338.026-00.476&Bright 8~and 24 \mum&PDR&...&0.29&28								&59		&...&...&14.0\\
        A16 &AGAL337.994-00.514&Bright 8~and 24 \mum&PDR&...&0.49&28								&267		&425&1.59&13.1\\  
        A17 &AGAL337.974-00.519&Bright 8~and 24 \mum&PDR&...&0.61&28								&242		&232&0.96&6.1\\ 
        A18	&AGAL337.939-00.532&Bright 8~and 24 \mum&PDR&...&0.45&28								&280		&517&1.84&17.8\\
        A19	&AGAL337.931-00.521&Bright 8~and 24 \mum&PDR&...&0.29&28								&96		&216&2.24&22.8\\
        A20 &AGAL337.934-00.507&Bright 8~and 24 \mum&PDR&...&0.80&28								&893		&443&0.50&10.1\\
		A21 	&AGAL337.889-00.489&Bright 8~and 24 \mum&PDR&...&0.36&28								&204		&169&0.83&25.3\\
		A22	&AGAL337.891-00.491&Bright 8~and 24 \mum&PDR&...&0.73&28								&397		&378&0.95&5.9\\
        A23 &AGAL337.916-00.477&Bright 8~and 24 \mum&PDR&16374-4701&0.68&28							&2462	&579&0.24&45.2\\ 
        A24 &AGAL337.922-00.456&Bright 8~and 24 \mum&PDR&...&1.16&28								&2699	&827&0.31&10.0\\ 
        A25	&AGAL337.927-00.432&Bright 8~and 24 \mum&PDR&16373-4658&1.62&28						&681		&1479&2.16&0.9\\
        
        B1&AGAL337.438-00.397&Bright 8 and 24 \mum&HII Region&16352-4721&0.86&24					&1221	&415		&0.34&11.1\\
        B2&AGAL337.406-00.402&4.5 \mum-e, and 24 \mum-p&Proto-stellar&16340-4732&0.66&19		&3673	&526	&0.14&73.8\\
        B3&AGAL337.392-00.396&Bright 8 and 24 \mum&HII Region&...&0.22&24							&99		&174&1.75&53.8\\
        B4&AGAL337.382-00.394&Bright 8 and 24 \mum&HII Region&...&0.28&24							&167		&138&0.82&44.0\\
        B5&AGAL337.371-00.399&Bright 8 and 24 \mum&HII Region&...&0.44&24							&301		&82&0.27&20.4\\
        
        B6&AGAL337.216-00.391&Dark&Pre-stellar&...&0.29&17										&353		&...&...&83.5\\
        B7&AGAL337.182-00.394&Dark&Pre-stellar&...&0.53&17										&794		&142&0.40&30.8\\
        B8&AGAL337.152-00.394&4.5 \mum-e, and 24 \mum-p &Proto-stellar&16351-4722&1.13&19		&2383	&...&...&9.5\\
        B9&AGAL337.139-00.382&4.5 \mum-e&Proto-stellar&...&1.01&19										&640		&1084&0.46&3.6\\
        B10&AGAL337.123-00.372&4.5 \mum-e&Proto-stellar&...&0.77&19										&503		&644&1.00&6.4\\
        B11&AGAL337.094-00.371&4.5 \mum-e&Proto-stellar&16352-4721&0.45&19								&313		&...&...&19.8\\
        B12&AGAL337.079-00.377&4.5 \mum-e&Proto-stellar&...&0.48&19										&495		&...&...&25.8\\

        C1&AGAL335.461-00.237&Dark&Pre-stellar&...&0.96&17										&1203	&490&0.41&7.8\\
        C2&AGAL335.441-00.237&Dark&Pre-stellar&16264-4841&0.65&17								&829		&785&0.95&17.4\\
        C3&AGAL335.427-00.241&4.5 \mum-e, and 24 \mum-p&Proto-stellar&...&0.80&19				&1562	&811&0.52&17.6\\
        C4 &AGAL335.402-00.241&24 \mum-p&Proto-stellar&...&0.45&19					&334		&218&0.65&21.2\\
        C5 &AGAL335.337-00.282&Dark&Pre-stellar&...&0.35&17										&410		&...&...&55.2\\
        C6 &AGAL335.306-00.284&Dark&Pre-stellar&...&0.45&17										&982		&...&...&62.2\\
        C7 &AGAL335.304-00.332&Dark&Pre-stellar&...&0.82&17										&1173	&...&...&12.3\\
        C8 &AGAL335.292-00.379&24 \mum-p&Proto-stellar&16263-4853&0.37&19				&200		&...&...&22.8\\
        C9 &AGAL335.289-00.281&Dark&Pre-stellar&...&0.26&17										&322		&...&...&105.6\\
        C10 &AGAL335.257-00.257&Dark&Pre-stellar&...&0.26&17									&99		&...&...&32.5\\
        C11 &AGAL335.249-00.304&Dark&Pre-stellar&...&1.12&17									&2002	&441&0.22&8.2\\
        C12 &AGAL335.236-00.322&Dark&Pre-stellar&...&0.33&17									&179		&125&0.70&28.7\\
        C13 &AGAL335.231-00.316&Dark&Pre-stellar&...&0.42&17									&469		&142&0.30&36.5\\
        C14 &AGAL335.221-00.344&Bright 8 and 24 \mum&HII Region&...&0.87&24						&771		&447&0.58&6.8\\
        C15 &AGAL335.212-00.377&Dark&Pre-stellar&...&0.38&17									&318		&131&0.41&33.5\\
        C16 &AGAL335.197-00.389&Bright 8 and 24 \mum&HII Region&16260-4858&0.52&24				&278		&...&...&11.4\\
        C17 &AGAL335.196-00.292&Bright 8 and 24 \mum&HII Region&...&0.31&24						&256		&...&...&49.7\\
        C18 &AGAL335.162-00.384&Dark&Pre-stellar&...&0.23&17									&177		&...&...&84.0\\
        C19 &AGAL335.122-00.414&Dark&Pre-stellar&...&0.26&17									&261		&...&...&85.9\\
        C20 &AGAL335.099-00.429&Dark&Pre-stellar&...&0.44&17									&603		&...&...&40.9\\
        C21 &AGAL335.061-00.427&4.5 \mum-e, and 24 \mum-p&Proto-stellar&16256-4905&1.17&19	&2277	&792&0.35&8.2\\
        C22 &AGAL335.059-00.401&Dark&Pre-stellar&...&0.42&17									&491		&208&0.42&38.2\\
   
          \hline
      \hline
    \end{tabular}
    }
  \end{center}
  \hspace{-5.3cm} $^b$ \footnotesize{Assumed dust temperature based on the work of Guzman et al. (in press.).}\\
   \hspace{-2cm} $^c$ \footnotesize{4.5 \mum-e refers to extended 4.5 \mum~emission, and 24 \mum-p refers to point like emission in this table.}
   
\end{table*}
\begin{table*}
\contcaption{} 
   \begin{center}
  {\scriptsize
    \begin{tabular}{l|cllcclcccr}
      \hline
      \hline
      Clump &\multicolumn{1}{|c}{ATLASGAL} & \multicolumn{1}{|c}{Infrared emission}&\multicolumn{1}{|c}{Classification}&\multicolumn{1}{|c}{IRAS} 
       & \multicolumn{1}{|c}{Diam.}& \multicolumn{1}{|c}{Dust} & \multicolumn{1}{|c}{m$_{dust}$} & \multicolumn{1}{|c}{m$_{vir}$}&\multicolumn{1}{|c}{$\alpha$}&\multicolumn{1}{|c}{$n$}\\
       & \multicolumn{1}{|c}{denomination}&& & &  &\multicolumn{1}{|c}{ Temp. }& & &&\multicolumn{1}{|c}{x 10$^4$}\\
       &             && & &\multicolumn{1}{|c}{ [pc]} &\multicolumn{1}{|c}{ [K]} & \multicolumn{1}{|c}{[M$_\odot$]}& \multicolumn{1}{|c}{[M$_\odot$]}&& \multicolumn{1}{|c}{[cm$^{-3}$]}\\
      \hline
      \hline

        D1&AGAL332.604-00.167&4.5 \mum-e, 24 \mum-p &Proto-stellar&16136-5038&0.96&19		&1596	&706&0.44&10.4\\
        D2&AGAL332.559-00.147&Bright 8~and 24 \mum&PDR&...&0.51&28									&495		&199&0.40&21.6\\
        D3&AGAL332.544-00.124&Bright 8~and 24 \mum&PDR& 16132-5039&0.96&28						&994		&261&0.26&6.5\\
        
        D4&AGAL332.531-00.159&Bright 8 and 24 \mum&PDR&...&1.22&28								&859		&666&0.78&2.7\\
        D5 &AGAL332.519-00.114&Bright 8~and 24 \mum&PDR&...&0.37&28								&110		&...&...&12.5\\
        D6 &AGAL332.506-00.119&Bright 8~and 24 \mum&PDR&...&0.30&28								&72		&...&...&15.4\\
        D7 &AGAL332.496-00.121&Bright 8~and 24 \mum&PDR&...&0.43&28								&186		&...&...&13.5\\
        D8 &AGAL332.469-00.131&Bright 8 and 24 \mum&HII Region&...&0.56&24							&399		&190&0.47&13.1\\
        D9 &AGAL332.442-00.139&Dark&Pre-stellar&...&0.53&17										&659		&266&0.40&25.6\\
        D10 &AGAL332.412-00.156&Dark&Pre-stellar&...&0.34&17										&230		&...&...&33.7\\
        D11 &AGAL332.399-00.151&24 \mum-p&Proto-stellar&...&0.40&19						&228		&50&0.22&20.5\\
        D12 &AGAL332.377-00.159&Dark&Pre-stellar&...&0.43&17										&277		&...&...&20.1\\
          D13 &AGAL332.352-00.116&Bright 8 and 24 \mum&Proto-stellar&16122-5028&0.64&19			&650		&221&0.34&14.3\\
        D14 &AGAL332.332-00.122&8 \mum~point source&Proto-stellar&...&0.27&19						&173		&81&0.47&50.8\\
        D15 &AGAL332.296-00.094&Bright 8 and 24 \mum&PDR&16119-5048&0.91&28						&1304	&511&0.39&10.0\\
        D16 &AGAL332.276-00.071&4.5 \mum-e, and 24 \mum-p&Proto-stellar&...&0.56&19			&472		&179&0.38&15.5\\
        D17 &AGAL332.254-00.056&Dark&Pre-stellar&...&0.98&17										&612		&416&0.68&3.8\\
        D18 &AGAL332.241-00.044&4.5 \mum-e, and 24 \mum-p&Proto-stellar&...&0.56&19			&1646	&502&0.31&54.1\\
        D19 &AGAL332.206-00.041&Dark&Pre-stellar&...&0.69&17										&751		&699&0.93&13.2\\
        D20 &AGAL332.204-00.024&Dark&Pre-stellar&...&0.50&17										&334		&155&0.93&15.4\\
        D21 &AGAL332.191-00.047&Dark&Pre-stellar&...&0.80&17										&970		&...&...&10.9\\

        E1 &AGAL332.467-00.522&4.5 \mum-e, and 24 \mum-p&Proto-stellar&16147-5100&0.86&19	&1700	&609&0.36&15.4\\
        E2 &AGAL332.411-00.471&8 and 24 \mum~emission&Proto-stellar&...&0.56&19					&532		&...&...&17.5\\
        E3 &AGAL332.409-00.504&Bright 8 and 24 \mum &PDR&16143-5101&0.81&28						&484		&416&0.86&5.3\\
        E4 &AGAL332.394-00.507&24 \mum-p&Proto-stellar&...&0.29&19						&109		&69&0.63&25.7\\
		E5 &AGAL332.351-00.436&Bright 8 and 24 \mum &HII Region&16137-5100&0.88&24				&798		&311&0.39&6.8\\
        E6 &AGAL332.336-00.574&Bright 8 and 24 \mum &PDR&...&0.70&28								&452		&141&0.31&7.6\\
        E7 &AGAL332.334-00.436&Dark&Pre-stellar&...&0.34&17										&451		&105&0.23&66.2\\
        E8 &AGAL332.326-00.454&Dark&Pre-stellar&...&0.65&17										&697		&134&0.19&14.6\\
        E9 &AGAL332.322-00.552&Bright 8 and 24 \mum &PDR&...&0.39&28								&264		&117&0.44&25.7\\
        E10 &AGAL332.312-00.556&Diffuse 8 and 24 \mum~emission&PDR&16141-5107&0.83&28				&660		&476&0.72&6.7\\
        E11 &AGAL332.281-00.547&4.5 \mum-e, and 24 \mum-p&Proto-stellar&...&0.77&19			&1251	&497&0.40&15.8\\
        E12 &AGAL332.252-00.539&Diffuse 8 and 24 \mum~emission&PDR&...&0.50&28						&375		&97&0.26&17.3\\
        E13 &AGAL332.226-00.536&4.5 \mum-e, and 24 \mum-p&Proto-stellar&...&0.83&19			&1096	&344&0.31&11.1\\    
		E14 &AGAL332.156-00.449&Bright 8 and 24 \mum &HII Region&16128-5109&0.93&24				&2557	&1057&0.41&18.3\\
		E15 &AGAL332.147-00.439&Bright 8 and 24 \mum &PDR&...&0.27&28								&240		&329&1.36&70.5\\
		E16 &AGAL332.144-00.469&Bright 8 and 24 \mum &PDR&...&0.86&28								&464		&...&...&4.2\\
		E17 &AGAL332.142-00.436&Bright 8 and 24 \mum &PDR&...&0.57&28								&353		&879&2.49&11.0\\
		E18 &AGAL332.141-00.466&Bright 8 and 24 \mum &PDR&...&0.59&28								&405		&821&2.02&11.4\\
		E19 &AGAL332.141-00.446&Bright 8 and 24 \mum &PDR&...&0.70&28								&813		&1736&2.14&13.7\\
		E20 & AGAL332.134-00.422&Bright 8 and 24 \mum &PDR&...&0.72&28							&376		&...&...&5.8\\
        E21 &AGAL332.094-00.421&Bright 8 and 24 \mum &HII Region&16124-5110&0.81&24				&1960	&491&0.25&21.3\\

         \hline
      \hline
    \end{tabular}
    }
  \end{center}
  \end{table*}

\end{document}